\documentclass[twocolumn,floatfix,pre]{revtex4}
\usepackage{graphicx}
\usepackage{amsfonts}
\usepackage{amssymb}
\usepackage{amsmath}

\begin{document}
\title{Lyapunov instability of rough hard-disk fluids}
\author{Jacobus A. van Meel}
\email{vanmeel@amolf.nl}
\affiliation{ FOM-Institute for Atomic and Molecular Physics, Kruislaan 407, 
1098 SJ Amsterdam, The Netherlands}
\author{Harald A. Posch}
\email{Harald.Posch@univie.ac.at}
\affiliation{ Faculty of Physics, Universit\"at Wien,
Boltzmanngasse 5, A-1090 Wien, Austria}

\date{\today}
\begin{abstract}
The dynamical instability of rough hard-disk fluids in two dimensions is characterized
through the Lyapunov spectrum and the Kolmogorov-Sinai entropy, $h_{KS}$, for a 
wide range of densities and moments of inertia $I$. For small $I$ the spectrum separates
into translation-dominated and  rotation-dominated parts. With increasing  $I$ the 
rotation-dominated part is gradually filled in at the expense of translation, until such
a separation becomes meaningless. At any density, the rate of phase-space mixing, given
by $h_{KS}$, becomes less and less effective the  more the rotation affects the dynamics.
However, the degree of dynamical chaos, measured by the maximum Lyapunov exponent, is only
enhanced  by the rotational degrees  of freedom for high-density gases, but is diminished
for lower densities. Surprisingly, no traces of  Lyapunov modes were found in the spectrum 
for larger moments of inertia. The spatial  localization of the perturbation vector 
associated with the maximum exponent however persists for any $I$.  
\end{abstract}
\maketitle


\section{Introduction}

The phase space trajectory of a many-body system with convex particles is Lyapunov
unstable \cite{posch_hoover_1988, gaspard_1998}, which means that   it is extremely
sensitive to small (infinitesimal) perturbations of the initial conditions. If all possible perturbation 
vectors are represented as an infinitesimal hypersphere centered on and 
co-moving with a phase point,  at later times this object is deformed under the 
action of the linearized flow. The main axes exponentially grow or shrink with time. The 
time-averaged rates are referred to as Lyapunov exponents, and the sorted set of exponents,
 $\{ \lambda_l, l=1 \dots D \}$, as Lyapunov spectrum. Here, $D$ is the phase-space dimension.
The relation of the Lyapunov spectrum to  more familiar 
fluid properties  has motivated much work over the last two decades. 
If, for example, the density of a soft-potential Weeks-Chandler-Anderson fluid
\cite{WCA} is isothermally increased from the fluid
to the solid phase, the Kolmogorov-Sinai (dynamical) entropy, 
which is equal to the sum of all positive Lyapunov exponents \cite{Pesin},
exhibits a maximum at a density  which is about 20 \% smaller than the liquid-line density
at the fluid-to-solid phase transition. This result, which has  been verified  for two- 
\cite{forster_posch_2005,toxvaerd_1983} and three-dimensional \cite{phh_1990,heyes_2006} 
systems, is somewhat surprising. It implies that  in soft-potential systems most efficient 
phase space mixing and information generation about the initial state does {\em not} occur at the  transition itself but  at a slightly smaller density. This sheds new light on our understanding of the onset
of phase transitions. Another example concerns stationary non-equilibrium systems with time-reversible equations of motion.  In this case,  the Lyapunov spectrum  may be  directly linked with the
transport coefficients and the rate of entropy production \cite{posch_hoover_1988,hoover_1999}.

Systematic studies of simple atomic fluids, such as hard-sphere and
soft-sphere models, have revealed some interesting features. In
particular, trajectory perturbations related to the largest (in absolute value) 
Lyapunov exponents are strongly localized in physical space 
\cite{hoover_posch_1998,FHPH_2004,TM_2003a}. This means that at  any instant of time
only a very small fraction of all particles contributes to the fastest dynamical events 
responsible for the largest rates of perturbation growth (or decay). This localization also
persists in the thermodynamic limit. A similar localization property for the 
maximum-exponent perturbations was found for one-dimensional distributed systems 
with space-time chaos \cite{pikovsky_1998}.
On the other hand, perturbation vectors associated with the
smallest (in absolute value) exponents may be delocalized and exhibit 
wave-like patterns in space. These so-called Lyapunov modes were first
observed for systems of hard dumbbells \cite{MPH_1998,MP_2002} and hard disks 
\cite{PH_2000,TM_2003b,EFPZ_2005} and are a consequence of the basic symmetries,
namely invariance with respect to time and space translations (depending on the boundary 
conditions). Due to exponent degeneracies, they are
recognized by a step-like appearance of the Lyapunov spectrum for
small (in absolute value) exponents. For soft-potential systems,
however, sophisticated Fourier transform techniques are required to demonstrate the
presence of Lyapunov modes \cite{forster_posch_2005, RADONS}.

The study of Lyapunov exponents has also been extended to simple
molecular systems such as soft \cite{Borzsak_1996,Kum_1998} and hard dumbbells 
\cite{MPH_1998,MP_2002} in two dimensions. In this case the dynamics is affected by
qualitatively different degrees of freedom, translation and rotation, which both have
a pronounced effect on the spectrum. 

In this paper we turn to a related model, which also incorporates the idea of 
the coupling of translational degrees of freedom to some internal energy,
which affects  the dynamics, namely rough hard disks in two dimensions.
The three-dimensional version of this model, rough hard spheres, was 
first introduced  by Bryan \cite{bryan:1894} in 1894, 
and was consecutively   treated by Pidduck \cite{pidduck:1922} and,
in particular,  by  Chapman and Cowling \cite{chapman:1953}, who worked out 
explicit formulas for the transport coefficients in terms of kinetic theory. 
It is an extension of the familiar smooth hard-sphere model, 
with  angular momentum added to each sphere due to roughness.
Roughness is introduced by the requirement that  a collision between two
particles reverses the relative surface velocity at the point of contact of the 
collision partners, exchanging both linear and angular momentum in the process.
This definition corresponds to the maximum possible roughness between
two particles. This model was extensively studied by Berne and co-workers
with respect to various models of molecular rotational relaxation \cite{Berne_I}.
They also investigated  the dynamical properties of models with partial roughness 
in between that of the smooth and the maximum rough sphere 
models \cite{Berne_II,Berne_III}, in particular the hydrodynamic long-time 
behavior of various correlation functions \cite{Berne_IV}. 

Here we are concerned with the two-dimensional version of this 
model with maximum roughness,  $N$ identical  rough hard disk in a box with
periodic boundaries. The paper is organized as follows:  
In Section \ref{section_model} we describe the model and derive the collision map for 
colliding particles and the respective linearized map for the
dynamics in tangent space.  All simulation results are presented in 
Section \ref{ergebnisse}. We conclude with a discussion of the results
in Sec. \ref{conclude}.

\section{Rough hard sphere and hard disk models}
\label{section_model}
%

\subsubsection*{Phase space dynamics}
We  consider the three-dimensional ($d=3)$ rough hard-sphere model  first.
It consists of  $N$ identical hard spheres  of 
diameter $\sigma$, mass $m$  and moment of inertia $I$, 
which are located at positions $\vec{q}_i$, and which move with (linear) velocities $\vec{v}_i$ 
and  rotate with angular velocities  $\vec{\omega}_i$, $i \in\{1,\cdots,N\}$.
Due to the isotropy of the spheres, the  particle orientation is not required in the 
following, and the state vector is given by
\begin{equation}
\vec{\Gamma} = \Big(\{ \vec{q}_i\}, \{ \vec{v}_i\}, \{\vec{\omega}_i\} \Big).
\end{equation}   
The phase space dynamics is characterized by force-free
''streaming'', interrupted by instantaneous pairwise collisions, at which
momentum and angular momentum are transferred between the
colliding particles, such that the relative surface velocity $\vec{g}$ at the 
point of contact of the two particles is reversed. Their  positions are not affected
by the collisions.

    The  equations of motion for the streaming between binary collisions
is written as a system of first-order differential equations,
\begin{equation}
\dot{\vec{\Gamma}} = \vec{F}(\vec{\Gamma}): \;\;  \Big(\{ \dot{\vec{q}}_i = \vec{v}_i \}, 
   \{\dot{\vec{v}}_i = \vec{0} \}, \{\dot{\vec{\omega}}_i = \vec{0}\} \Big),
 \label{streaming}
 \end{equation}           
where $i=1,\dots, N$. It has an  obvious solution. 

     At a collision between two particles $i$ and $j$,  the collision map 
\begin{equation}
\vec{\Gamma}\,' = \vec{M}(\vec{\Gamma})
\label{collmap}
\end{equation}
 is introduced, which  relates the 
pre-collision state  $\vec{\Gamma}$ to the post-collision state $\vec{\Gamma}\,'$.
With the following definitions,
\begin{equation}
\vec{q} =  \vec{q}_j - \vec{q}_i; \;\;\;
\vec{n} = \frac{1}{\sigma} \vec{q}; \;\;\;
\vec{v} = \vec{v}_j - \vec{v}_i; \;\;\;
\vec{\Omega} = \vec{\omega}_j + \vec{\omega}_i, 
\label{defps}
 \end{equation}
the relative surface velocity at the point of contact of $i$ and $j$ becomes 
\begin{equation}
\vec{g} = \vec{v} + \frac{\sigma}{2}\, \vec{n} \times \vec{\Omega}.
\label{defg}
\end{equation}
Here, $\vec{n}$ denotes a unit vector along the line of centers from $i$ to
$j$ at the instant of collision, and $\vec{v}$ is the respective relative velocity.
Maximum roughness requires that 
\begin{equation}
      \vec{g}\,' = - \vec{g},
\end{equation}      
where the prime, here and below,  refers to the state immediately after the collision. 
Together with the conservation laws for energy, linear momentum and angular momentum, 
that part of the collision map relevant for the collision between particles $i$ and $j$
becomes \cite{chapman:1953}
\begin{eqnarray}
\vec{q}_i\,' &=& \vec{q}_i, \nonumber\\
\vec{q}_j\,' &=& \vec{q}_j, \nonumber \\
\vec{v}_i\,' &=& \vec{v}_i + \gamma \vec{g} + \beta \, \vec{n} 
( \vec{n} \cdot \vec{v} ), \nonumber \\
\vec{v}_j\,' &=& \vec{v}_j - \gamma \vec{g} - \beta \, \vec{n} 
( \vec{n} \cdot \vec{v} ), \\
\vec{\omega}_i\,' &=& \vec{\omega}_i + (2 \beta/\sigma) \,\vec{n} \times \vec{g}, \nonumber\\
\vec{\omega}_j\,' &=& \vec{\omega}_j + (2 \beta/\sigma) \, \vec{n} \times \vec{g}. \nonumber
\end{eqnarray}
Here, $\beta$ and $\gamma$ are constants, which depend on the moment of
inertia $I$ around a diameter of the sphere:
\begin{equation}
\kappa = \frac{ 4 I }{m \sigma^2}; \qquad
\gamma = \frac{\kappa}{\kappa + 1}; \qquad
\beta = \frac{1}{\kappa + 1}.
\end{equation}
The dimensionless parameter $\gamma$  controls  the coupling between 
translational and rotational degrees of freedom. In the limit $I \rightarrow 0$, the
translational and rotational degrees of freedom decouple and the 
rough hard sphere model reduces to the conventional smooth 
hard-sphere model without roughness \cite{DPH_1996}, if all angular
velocity vectors are discarded.

\subsubsection*{Tangent space dynamics}
A perturbed trajectory is separated from the reference trajectory by
an offset vector
\begin{equation}
\vec{\delta\Gamma} = (\{ \vec{\delta q}_i\}, \{ \vec{\delta v}_i\}, \{\vec{\delta \omega}_i\} ),
\end{equation}   
which evolves during the streaming phase according to the linearized equations of motion
\begin{equation}
\dot{\vec{\delta \Gamma}} = \frac{\partial\vec{F}}{\partial \vec{\Gamma}} \cdot
      \vec{\delta \Gamma}:  \Big(\{ \dot{\vec{\delta q}}_i = \vec{\delta v}_i \}, \{\dot{\vec{\delta v}}_i = \vec{0} \}, 
       \{\dot{\vec{\delta \omega}}_i = \vec{0} \} \Big),
 \label{tpstreaming}
 \end{equation}           
$i=1,\cdots,N$. It is trivially solved.

    The linearization of the collision map (\ref{collmap}) is obtained from Ref. \cite{DPH_1996}:
\begin{equation}
\vec{\delta \Gamma}\,' = \frac{\partial\vec{M}}{\partial \vec{\Gamma}} \cdot \vec{\delta\Gamma}
   + \left[   \frac{\partial\vec{M}}{\partial \vec{\Gamma}} \cdot \vec{F}(\vec{\Gamma}) -
      \vec{F}\left(\vec{M}(\vec{\Gamma})   \right) \right]  \delta \tau_c.
 \label{lcm}     
\end{equation}    
Here, $\delta \tau_c$ is the (infinitesimal) time shift between the collision of the 
reference trajectory and of the perturbed trajectory, which may be positive or negative.
Similarly, we denote by $\delta \vec{q}_c $ the shift in configuration space  of the
collision points of these two trajectories. If,  as before, $i$ and $j$
are the colliding particles, these quantities are computed from \cite{DP_1997}
\begin{equation}
\delta \tau_c = - \frac{ \delta \vec{q} \cdot \vec{n}}{\vec{v} \cdot \vec{n}}; \qquad
\delta \vec{q}_c =   \delta \vec{q} + \vec{v}\, \delta \tau_c,
\end{equation}
where we use a notation for the perturbed quantities which is analogous 
to the previous notation in phase space (see Eq. (\ref{defps})): 
\begin{equation}
\delta \vec{q} = \delta \vec{q}_j - \delta \vec{q}_i; \;\;\;\;
\delta \vec{v} = \delta \vec{v}_j - \delta \vec{v}_i; \;\;\;\;
\delta \vec{\Omega} = \delta \vec{\omega}_j + \delta \vec{\omega}_i.
\end{equation}
Since $\vec{\delta q}_c = \sigma \vec{\delta n}$, we also find from Eq. (\ref{defg}):
\begin{equation}
\delta \vec{g} = \delta \vec{v} + \frac{1}{2} \Big[ \delta \vec{q}_c
\times \vec{\Omega} + \vec{q} \times \delta \vec{\Omega} \Big].
\end{equation}
With this notation, we obtain for that part of the linearized map (\ref{lcm}) belonging
to the collision of $i$ and $j$:  
\begin{eqnarray}
\delta \vec{q}_i\,' &=& \delta \vec{q}_i - \Big[ \gamma \vec{g} + \frac{\beta}{\sigma^2} \vec{q}
( \vec{q} \cdot \vec{v} ) \Big] \delta \tau_c \,,  \nonumber \\
\delta \vec{q}_j\,' &=& \delta \vec{q}_j + \Big[ \gamma \vec{g} + \frac{\beta}{\sigma^2} \vec{q}
( \vec{q} \cdot \vec{v} ) \Big] \delta \tau_c  \,,\nonumber \\
\delta \vec{v}_i\,' &=& \delta \vec{v}_i + \gamma \, \delta \vec{g} \nonumber \\ 
 & & + \frac{\beta}{\sigma^2}
\Big[ \delta \vec{q}_c ( \vec{q} \cdot \vec{v} ) + \vec{q} ( \vec{v} \cdot
\delta \vec{q}_c ) + \vec{q} ( \vec{q} \cdot \delta \vec{v} ) \Big] \,, \nonumber \\
\delta \vec{v}_j\,' &=& \delta \vec{v}_j - \gamma \, \delta \vec{g}  \nonumber \\
& & -\frac{\beta}{\sigma^2} \Big[ \delta \vec{q}_c ( \vec{q} \cdot \vec{v} ) + \vec{q} ( \vec{v} \cdot
\delta \vec{q}_c ) + \vec{q} ( \vec{q} \cdot \delta \vec{v} ) \Big]  \,,\nonumber  \\
\delta \vec{\omega}_i\,' &=& \delta \vec{\omega}_i + \frac{2\beta}{\sigma^2} \Big[ \delta
\vec{q}_c \times \vec{g} + \vec{q} \times \delta \vec{g} \Big]  \,, \nonumber \\
\delta \vec{\omega}_j\,' &=& \delta \vec{\omega}_j + \frac{2\beta}{\sigma^2} \Big[ \delta
\vec{q}_c \times \vec{g} + \vec{q} \times \delta \vec{g} \Big]  \label{lcmcomp}.
\end{eqnarray}
If the  moment of inertia  vanishes ($\gamma \to 0$ and $\beta \to 1$), the
linearized collision map of the smooth hard sphere fluid is recovered \cite{DP_1997},
if the angular velocity perturbations are discarded.

\subsubsection*{Computer simulations of hard disks in two dimensions}

For the remainder of this work we restrict ourselves to the case of planar rough disks
on the $xy$ plane, $d = 2$. All the equations above remain valid in this case,
if  all position and velocity vectors are placed in the $xy$ plane, 
and all angular velocity vectors $\vec{\omega}_i$
are perpendicular to this plane with a single non-vanishing $z$ component $\omega_i$.
Discarding all superfluous components, we are left with $D = 5N$ components for the state vector 
$\vec{\Gamma}$ and for any perturbation vector $\vec{\delta \Gamma}$.

Depending on the mass distribution of the disks and, hence, their moment of inertia $I$ with 
respect to a perpendicular axis through the center, the coupling parameter $\kappa$ may take 
values between zero and one:
$\kappa = 0$ applies, if all the mass is located in the 
center,  $\kappa = 1/2$ corresponds to a uniform mass distribution, and $\kappa = 1$
is obtained,  if all the mass is concentrated on the perimeter of the disk.

For our numerical work we use reduced units, for which the disk
diameter $\sigma$, its mass $m$, and the Boltzmann constant
$k$ are set to unity. The time-averaged translational kinetic energy per particle,
$\langle K \rangle /N = \sum_i m \langle \vec{v}_i \cdot \vec{v}_i \rangle/(2N)$
is taken as the unit of energy.  Thus the temperature is  one, $T = 1$, 
as in our previous work on smooth hard disks, which facilitates comparison
\cite{DPH_1996}.  We have
ascertained that the translational and rotational temperatures agree. The 
unit of time is $(m \sigma^2 N / \langle  K \rangle)^{1/2}$. Lyapunov exponents and 
the Kolmorov-Sinai entropy are measured in units of $\sqrt{k T / m \sigma^2}$.
The number density is defined as $\rho = N \sigma^2/ V$, where $N$ is the total number of
disks, and $V $ denotes the area of  the simulation box with extensions $L_x, L_y$. 
For fluid phases we take  a square box, $L_x = L_y$,  where the particles are 
initially put on  a square lattice with random velocities and random angular velocities. 
For solid phases the aspect ratio
$L_y/L_x = \sqrt{3}/2$ is used, which is compatible with the triangular close-packed
lattice and does not differ much  from a square. By initially putting the particles on a 
triangular  lattice with random velocities and angular velocities, high density systems 
up to the close-packed density $\rho_0 = 1.1547$ may be studied in this way.   
Periodic boundary conditions are used throughout. The only free parameters 
are the number density $\rho$ and the moment of inertia $I$ (respective the
coupling parameter $\kappa = 4 I$).

To evolve the system, an event-driven algorithm
similar to that of  Rapaport  \cite{Rapaport} was used. Proper care
was taken to avoid missing glancing collisions. The conservation of
energy and  linear momentum was carefully verified. The total angular
momentum is not conserved due to the periodic boundaries. 
For the computation of the $5N$ Lyapunov exponents, the classical
algorithm of Benettin et al. \cite{Benettin} and Shimada et al. \cite{Shimada},
properly modified for the instantaneous elastic rough collisions encountered here, 
was used. Simultaneously with the reference trajectory  $\vec{\Gamma}(t)$, $5N$ replica of 
perturbation vectors, $\vec{\delta \Gamma}^{(l)}(t), \; l=1,\cdots, 5N$, with orthonormal
initial conditions were evolved and periodically   re-orthonormalized with a
Gram-Schmidt procedure \cite{DPH_1996,DP_1997}.

\section{Results}
\subsection*{Lyapunov spectra}
\label{ergebnisse}

In this section we discuss the Lyapunov spectra of rough disk fluids and solids.
A full spectrum consists of the
ordered set of Lyapunov exponents, $\lambda_1 \ge \lambda_2 \ge \cdots \ge \lambda_{D}$,
where $D$ is the phase space dimension. For a symplectic system as in our case,
the Lyapunov exponents always appear in pairs with a vanishing
pair sum. Therefore, the Lyapunov spectrum consists of a positive
and a negative branch and is symmetric,  $\lambda_l = - \lambda_{D + 1 - l}$,
if plotted as a function of the index $l$. In the absence of magnetic fields, this so-called
''conjugate pairing symmetry'' may simply be viewed as a consequence of the
time-reversibility of the dynamical equations. Therefore, we may restrict ourselves
to the positive branch of the spectrum, $1\le l \le D/2$, for which $\lambda_l \ge 0$,
which considerably reduces the computational effort. 
Examples for full spectra will be given below (see the upper panel of Fig. \ref{Fig_8}).

Another important property concerns the vanishing exponents, which are a consequence
of the fundamental continuous symmetries which leave the Lagrangian and, hence, 
the motion equations invariant. According to N\"other's theorem, each such symmetry 
corresponds to a constant of the motion \cite{Scheck} and, in addition, generates two 
symplectic-conjugate vector fields in phase space, along which perturbations do not
stretch or shrink and, therefore, give rise to two vanishing exponents \cite{gaspard_1998}.
For example, invariance with 
respect to time translation gives rise to energy conservation and to two vanishing exponents.
Similarly, invariance with respect to uniform translation in space gives rise to momentum conservation
and four more vanishing exponents. However, isotropy of space and, hence, angular momentum
conservation applies locally for each collision, but not globally due to the periodic 
boundary conditions.  Altogether only 6 Lyapunov exponents vanish in our case, 
as is also confirmed by the simulations.

\begin{figure}[ht]
\center
\includegraphics[width=0.45\textwidth]{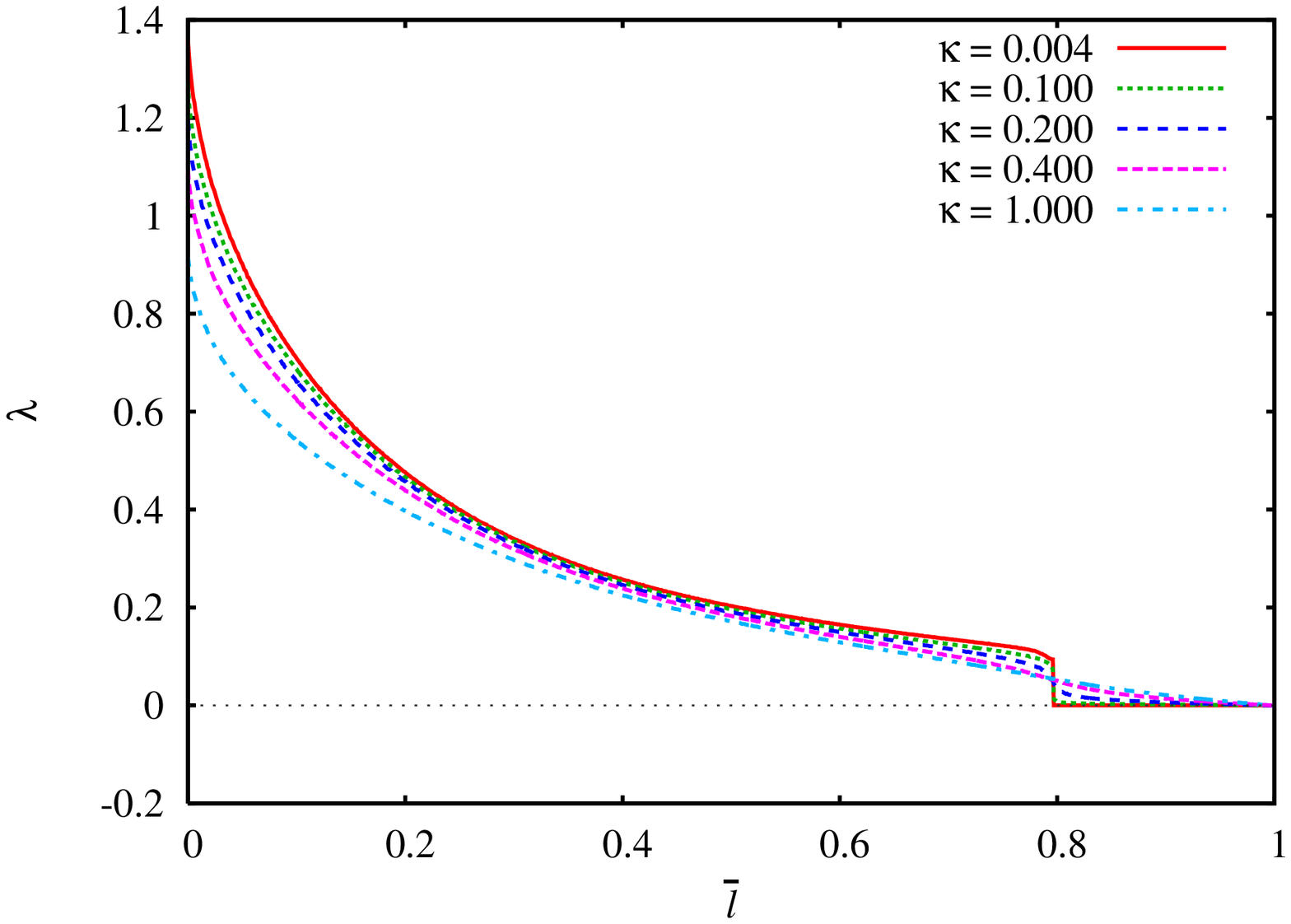}
\includegraphics[width=0.45\textwidth]{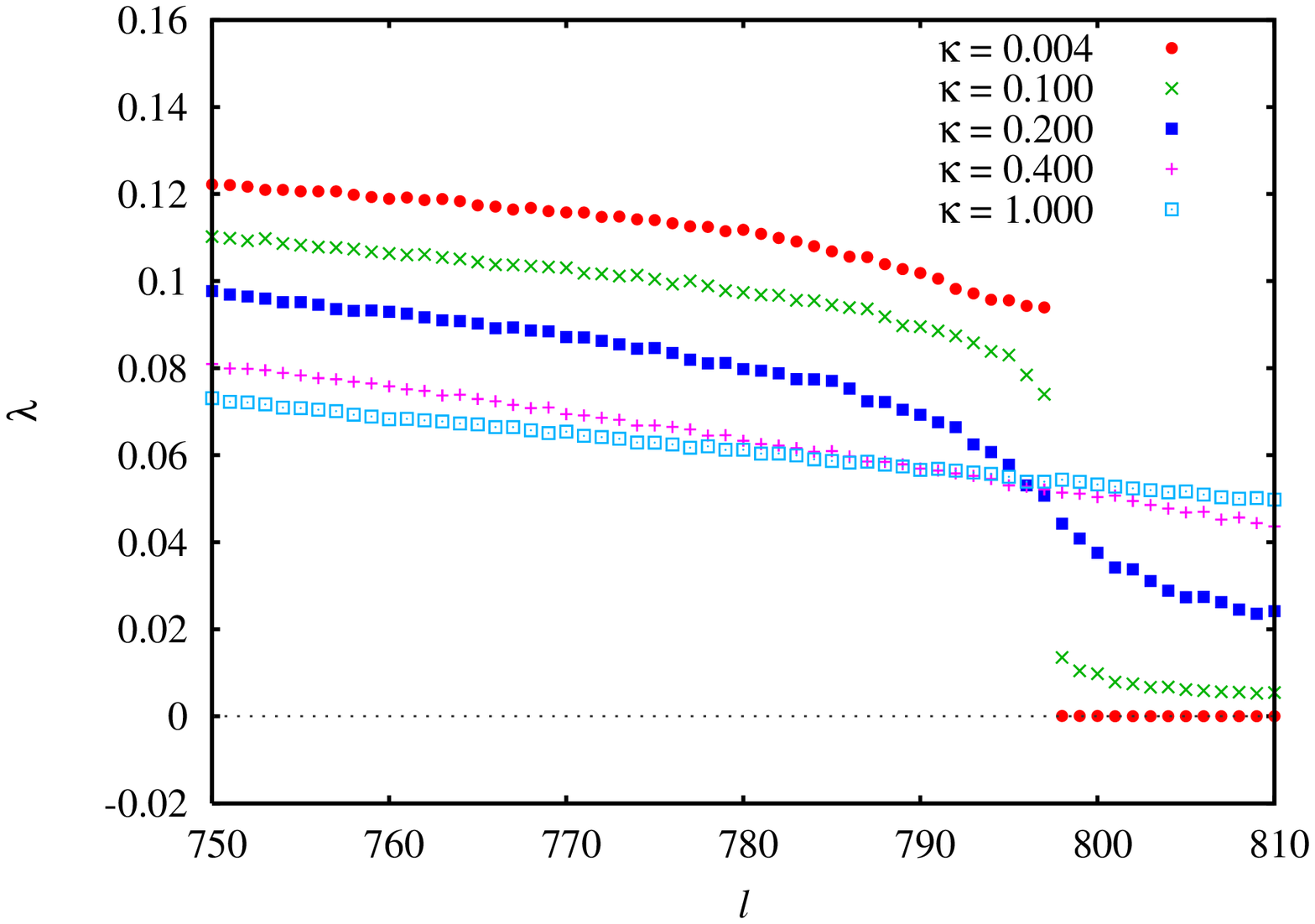}
\caption{(Color online) Lyapunov spectra for a system of $N = 400$ rough hard disks at a
low density $\rho = 0.1$. The various curves are for different moments of inertia $I$, where the keys  refer to the coupling  parameter $\kappa = 4I$. The uniform mass distribution 
corresponds to $\kappa = 0.5$.
Top panel: Positive branches of the specta. The reduced index $\bar{l}$ is used on the abscissa.
Bottom: Magnification of the transition region between translation and rotation dominated exponents.
The un-normalized index $l$ is used on the abscissa. }
\label{Fig_1tb}
\end{figure}

\begin{figure}[ht]
\center
\includegraphics[width=0.45\textwidth]{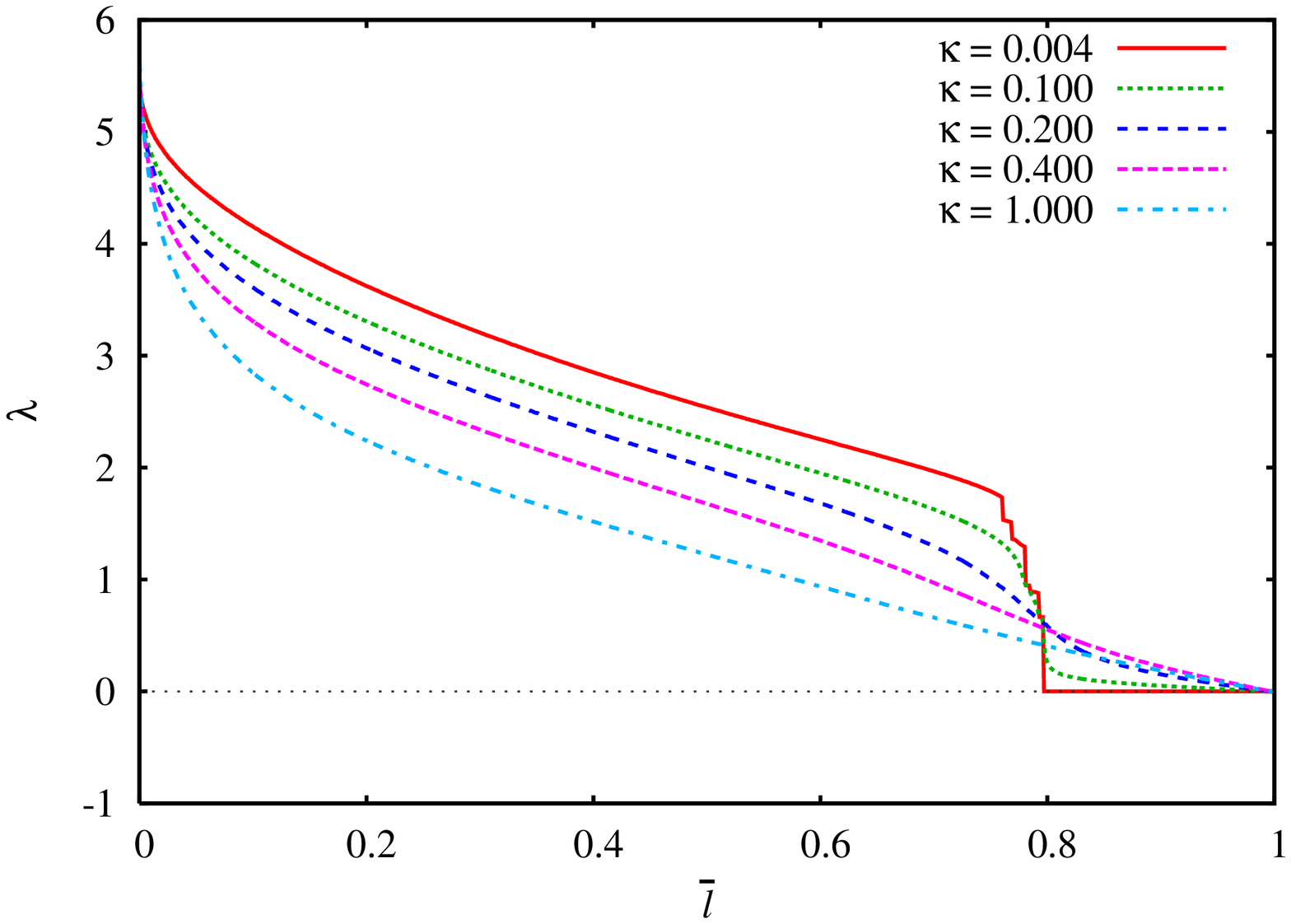}
\includegraphics[width=0.45\textwidth]{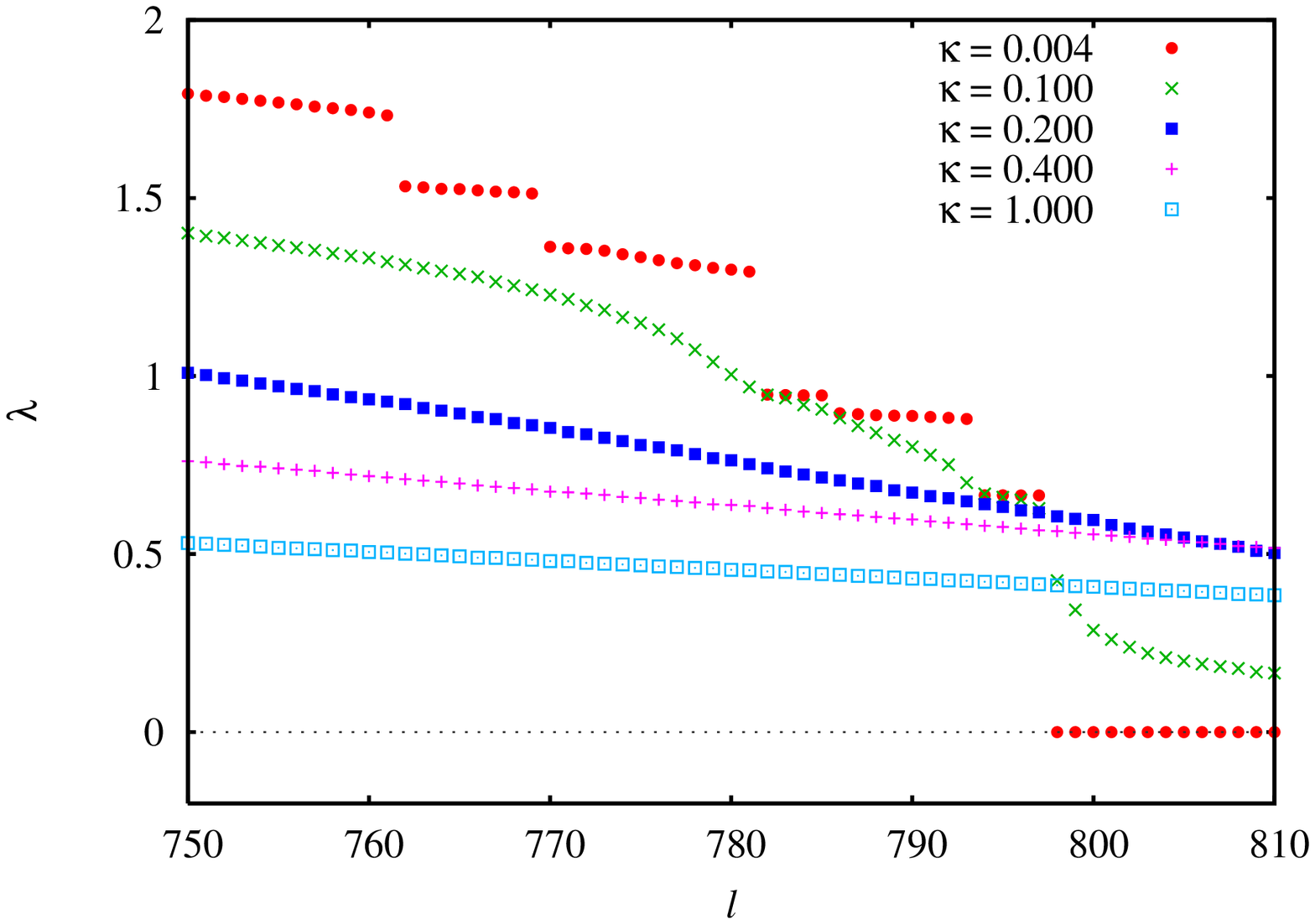}
\caption{(Color online) Lyapunov spectra for a system of $N = 400$ rough hard disks at a 
moderately high density $\rho = 0.7$ (top). The various curves are for different moments 
of inertia, where the keys specify  the coupling parameter $\kappa = 4I$.
The uniform mass distribution corresponds to  $\kappa = 0.5$.
Top panel: Positive branches of the specta. The reduced index $\bar{l}$ is used on the abscissa.
Bottom panel: Magnification of the transition region between translation and rotation dominated exponents. The un-normalized index $l$ is used on the abscissa.
Reduced units and periodic boundaries are  used as explained in the main text.}
\label{Fig_2tb}
\end{figure}

Although a spectrum only consists of discrete points indexed by the integer $l$ --
or by the reduced index $\bar{l} = 2l/D$ - most of the time in the figures below it is represented 
by a smooth line drawn through these points to enhance the clarity. 

       To assess the influence of translation-rotation coupling on the Lyapunov spectra,
we show in Fig. \ref{Fig_1tb} results for a rather
dilute gas, $\rho = 0.1$, of $N = 400$ rough disks at a temperature $T=1$. 
The various curves  belong to different moments of inertia $I$ and are specified by their 
coupling parameters $\kappa = 4I$. The system is fairly large, and the results are
close to the thermodynamic limit \cite{DPH_1996}. The phase space has $D = 2000$ dimensions.
In the top panel of Fig.  \ref{Fig_1tb}  the positive branches of the full spectra are shown, where  the normalized index  $\bar{l} = l/(D/2)$ is used ($1 \le  l \le D/2 = 1000$) on the abscissa.
Most noticeable is the transition region near $\bar{l}_0 = (2N-3)/(D/2) = 0.797$, which 
separates the specta 
into a translation-dominated regime for $\bar{l} \le \bar{l}_0$ and a  rotation-dominated regime
for $\bar{l}_0 < \bar{l} \le (5N-6)/D)$. A magnification of the transition region is shown in the lower
panel  of Fig. \ref{Fig_1tb}, where the un-normalized index $l$ is used on the horizontal
axis. Analogous spectra for a  rather dense gas, $\rho = 0.7$, are shown in Fig. \ref{Fig_2tb}.

For very small $\kappa$  translation and rotation 
are effectively decoupled and the dynamics is almost identical to that of a smooth hard-disk
gas. For 400 particles  the positive exponents $\lambda_1, \cdots, \lambda_{797}$
agree with the positive exponents of the smooth hard disks and, hence, are translation dominated, 
whereas the (very small but positive) exponents $\lambda_{798}, \cdots \lambda_{997}$ are due
to the angular velocity perturbations and are only present in the rough-disk case. The
three remaining exponents, $\lambda_{998}, \cdots, \lambda_{1000}$, vanish due to the 
conserved quantities. Note that this is only the positive branch of the full spectrum.

\begin{figure}[ht]
\center
\includegraphics[width=0.45\textwidth]{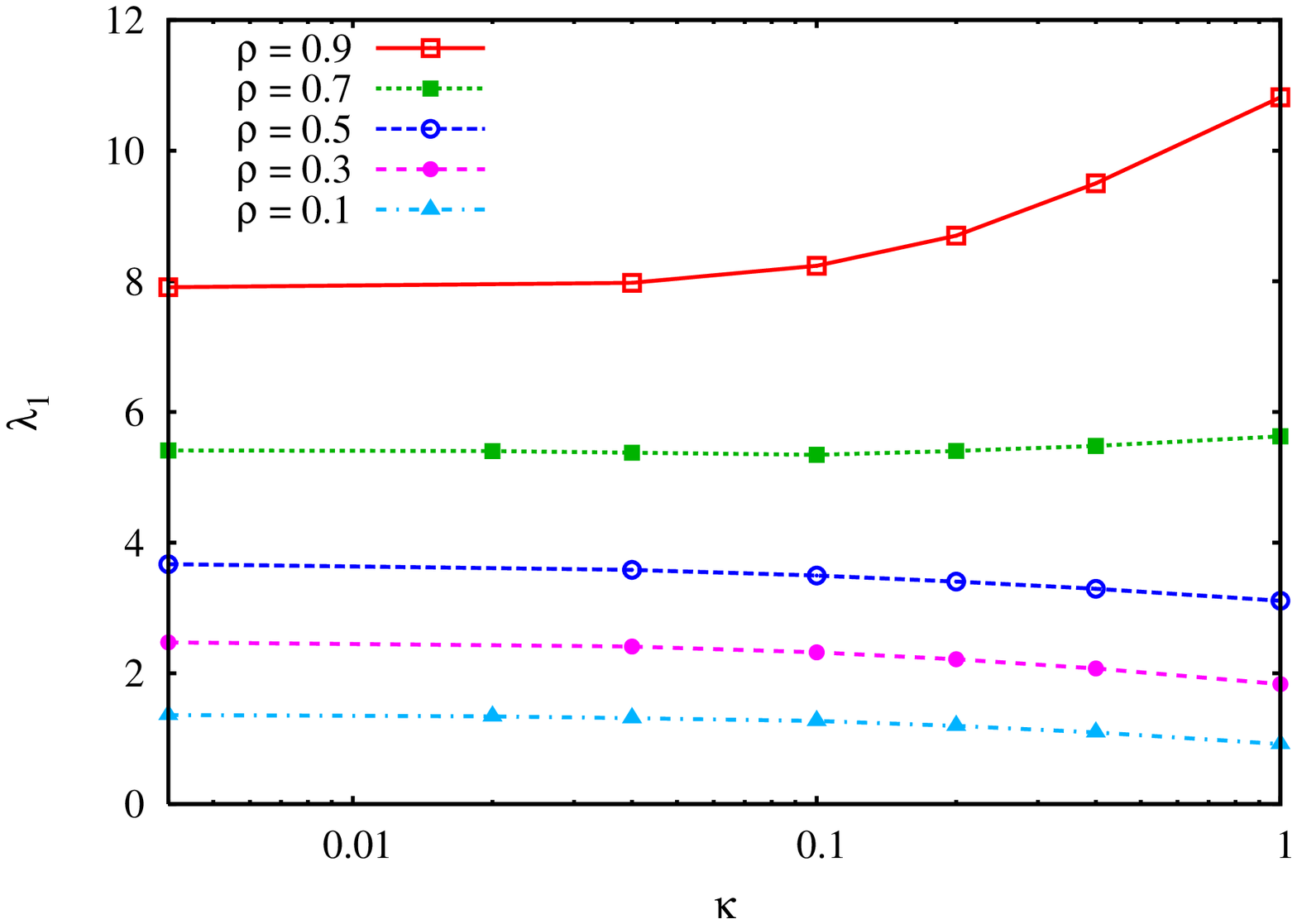}
\includegraphics[width=0.45\textwidth]{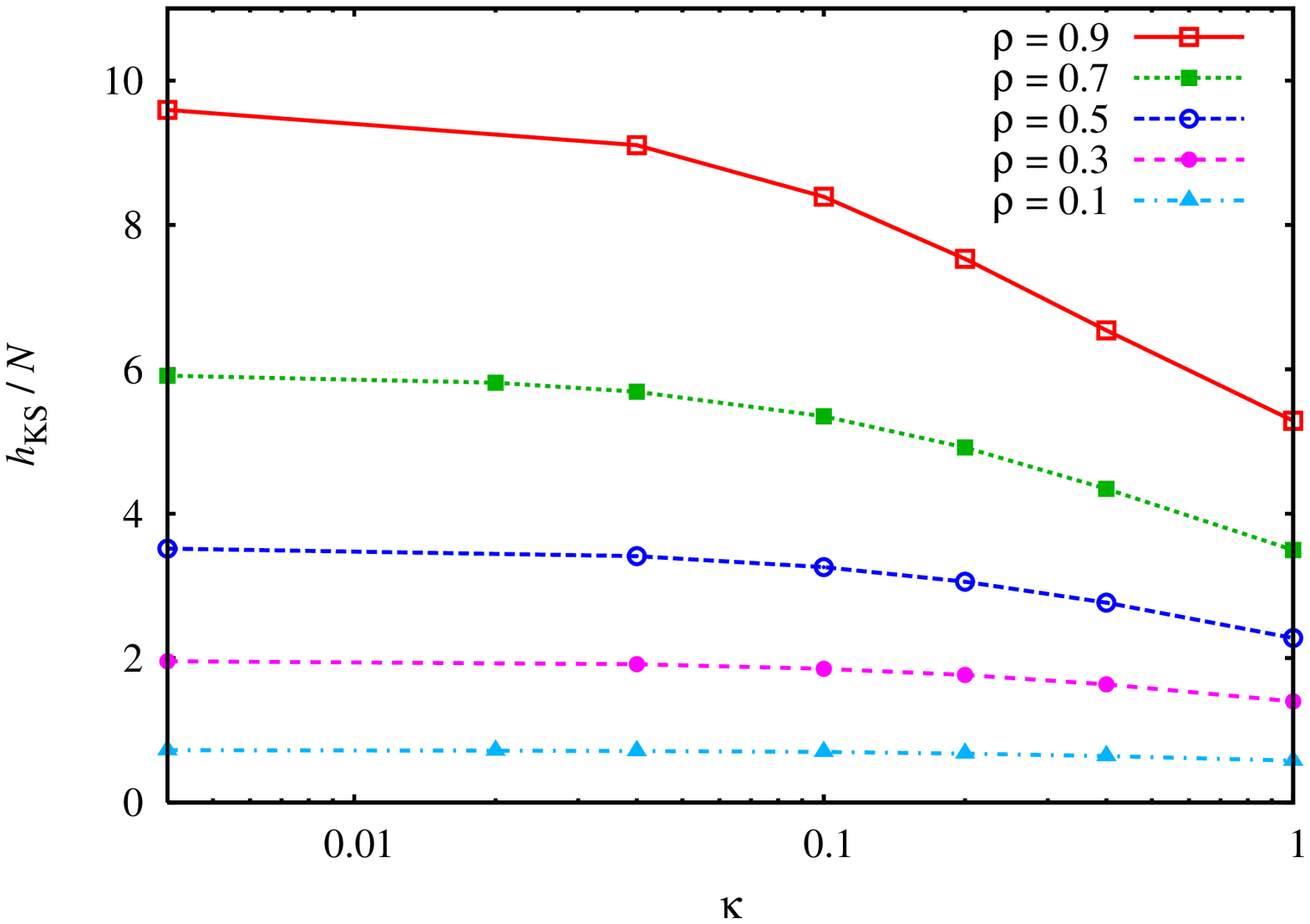}
\caption{(Color online) Dependence of the maximum Lyapunov exponent
$\lambda_1$ (upper panel) and of the Kolmogorov-Sinai entropy per particle,
$h_{KS}/N$, (lower panel) on the coupling parameter $\kappa$ for various densities
as indicated by the keys. The rough disk fluid contains N=400 particles and has 
(translational) temperature $kT = 1$.  Reduced units are used throughout.  } 
\label{Fig_3tb}
\end{figure}
The maximum Lyapunov exponent $\lambda_1$ is generally taken as in indicator and a
measure for dynamical chaos. Similarly,  the Kolmogorov-Sinai (or dynamical) entropy 
$h_{KS}$ is a measure of phase-space mixing \cite{DP_1997a}.
Due to the exponential instability, a number of initially close phase points are eventually uniformly
distributed over the energy surface. The characteristic time for this mixing process is 
the mixing time $1/h_{KS}$ \cite{Krylov,Zaslavsky}. Since, according to Pesin \cite{Pesin},
$h_{KS}$ is equal to the sum of the positive exponents, it is directly accessible through the
Lyapunov spectrum. In the upper panel of Fig. \ref{Fig_3tb} the 
maximum exponent as a function of $\kappa$ is shown for various densities. 
An analogous plot for the KS-entropy per particle, $h_{KS}/N$, is provided 
in the lower panel of the same figure. If $\kappa$ is increased, $\lambda_1$ 
decreases - weakly - for lower densities, $\rho < 0.7$, and increases for large densities. The KS-entropy
always decreases with $\kappa$, even for large densities. This means that
mixing becomes less effective the more the rotational degrees of freedom affect the 
translational dynamics. 
Similarly, chaos is  slightly reduced  with increasing $\kappa$, at least for low-density gases.

\begin{figure}[ht]
\center
\includegraphics[width=0.45\textwidth]{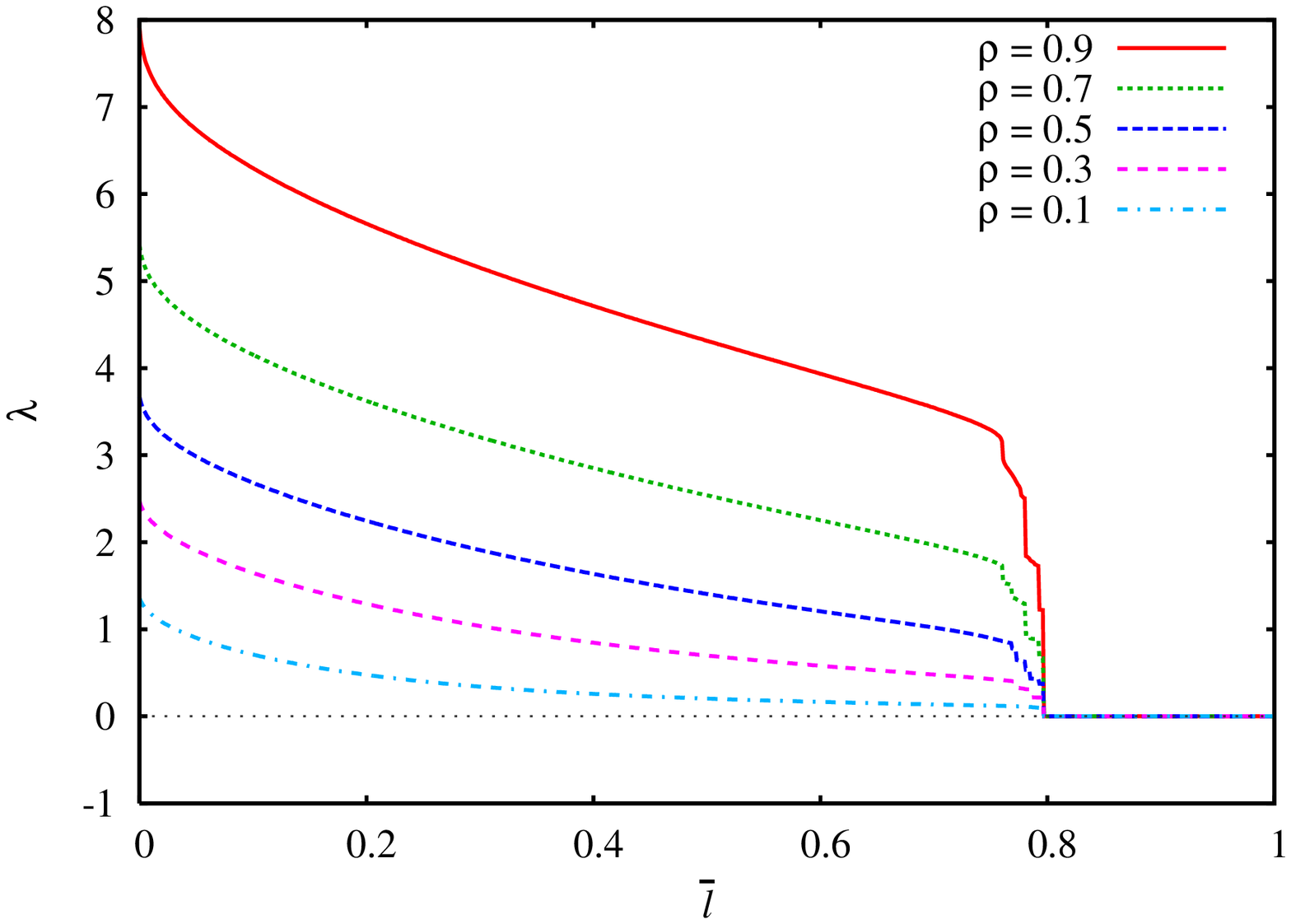}
\includegraphics[width=0.45\textwidth]{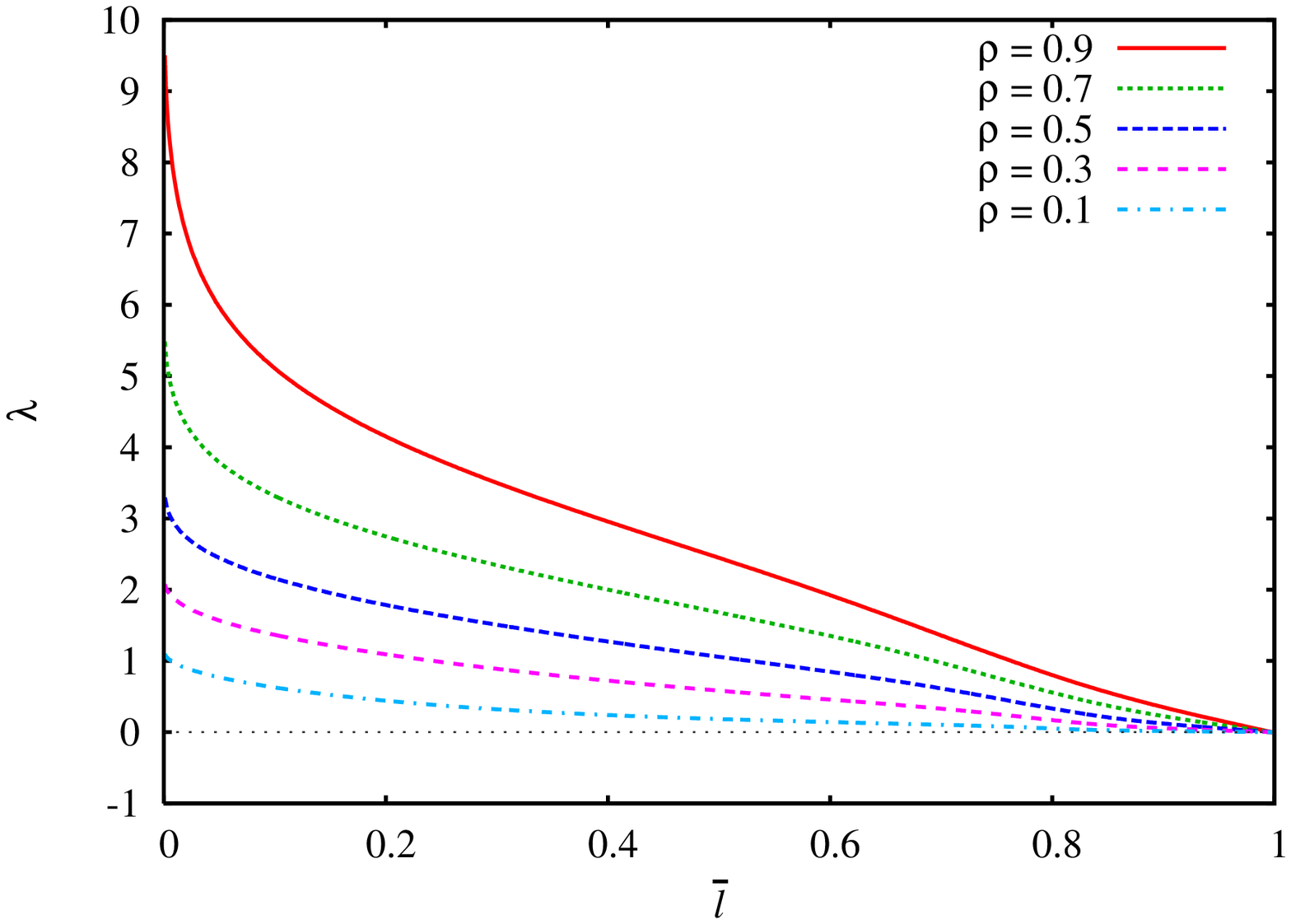}
\caption{(Color online) Lyapunov spectra (positive branch only) of  400
       rough hard disks with very small moment of inertia ($\kappa = 4I = 0.004$, upper panel)
       and with a large moment of inertia ($\kappa = 0.4$, lower panel) for various densities as specified 
       by the key. The reduced index $\bar{l}$ is used on the abscissa. }
\label{Fig_4tb}
\end{figure}
\begin{figure}[ht]
\center
\includegraphics[width=0.45\textwidth]{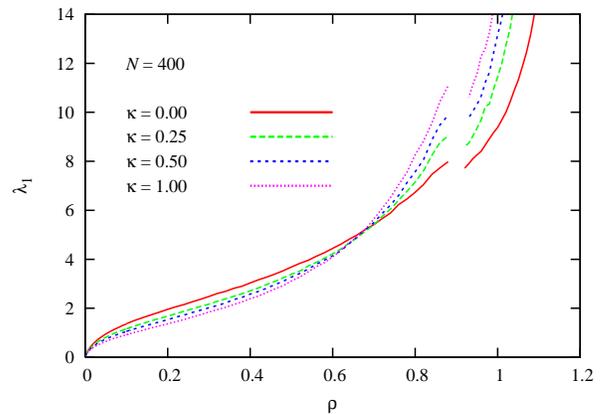}
\caption{(Color online) Density dependence of the maximum exponent of  400 rough hard disks  for
various moments of inertia corresponding to the specified values of $\kappa$.  }
\label{Fig_5}
\end{figure}

To demonstrate the density dependence, we show in the upper panel of Fig. \ref{Fig_4tb} 
Lyapunov spectra for various densities of a 400-particle gas of hard disks with a 
very small moment of inertia  for which $\kappa = 0.004$. Analogous spectra with a 
large moment of inertia corresponding to $\kappa = 0.4$ are provided in the lower panel
of the same figure. All exponents increase with the density, in particular $\lambda_1$, as
is shown in more detail in Fig. \ref{Fig_5}. where the maximum exponent  - for various $\kappa$ -
is plotted as a function of $\rho$.  In some sense, $\lambda_1$ behaves similar to the 
potential-generated contribution to the pressure $P$ \cite{FMP_2004}. For low densities, 
both $(P / \rho kT) - 1$ and $\lambda_1$ are proportional to the single particle collision 
frequency $\nu_2$. Fig. \ref{Fig_5} resembles the respective phase diagram for the pressure.
The conspicuous gap in the spectra marks the two-phase region for the fluid-solid
transition. As mentioned in the previous section, the data beyond the solid line were obtained with a non-square simulation box (aspect ratio $\sqrt{3}/2$), the data below the fluid line with a square
simulation box. This choice of aspect ratios merely facilitates the setting up of the initial
conditions for the solid-state simulations and does not have any significance for the results.
The gap disappears completely, if $\lambda_1$ is plotted as a function of the
single-particle collision frequency $\nu_2$ (not shown), which is easily obtained from the simulation.
This has been noted already for the smooth hard-disk system ($\kappa = 0$)
\cite{DPH_1996} and is also true for all $\kappa > 0$. This means that the statistical 
distributions for the parameters characterizing the collisions (such as the impact parameter)
do not noticeably differ for the disordered fluid and the coexisting crystal. 

Based on kinetic theory, a density expansion for $\lambda_1$   of the smooth hard disk model
($\kappa = 0$) becomes 
\begin{equation}
      \lambda_1 = A \nu_2 \left[ - \ln \rho - B + {\cal O} (1 / \ln \rho) \right],
      \label{vanZon}
\end{equation}      
where 
\begin{equation}
       \nu_2 =   2 \pi^{1/2} \rho  \sigma  \left( kT/m \right) g(\sigma)
\end{equation}        
is the single-particle collision frequency. The pair distribution function at
contact, $g(\sigma)$, converges to  unity in the low-density limit.  Estimates
for the constants $A$ and $B$ have been computed by
van Zon and van Beijeren,  $A = 1.473, B = 2.48 $ \cite{vZ1,vZ2}, which
represent the numerical data well for very small densities $\rho <  10^{-3}$.
Expressions such as Eq. \ref{vanZon} with different constants are expected to hold 
also for the rough disks when  $\kappa > 0$.       
 
\begin{figure}[b]
\center
\includegraphics[width=0.45\textwidth]{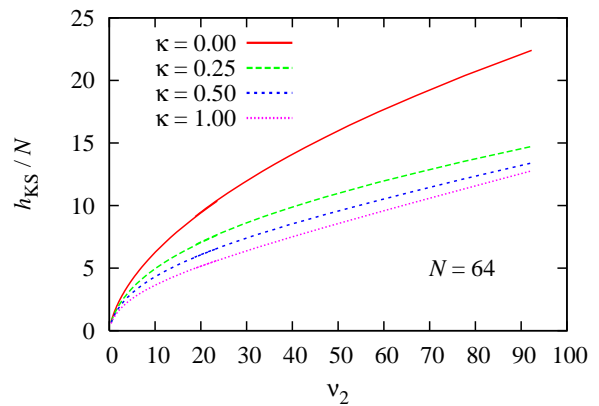}
\caption{(Color online) Dependence of the Kolmogorov-Sinai entropy $h_{KS}$ on the single-particle
collision frequency  $\nu_2$ of $N = 64$ rough disks for various $\kappa$ as indicated by the
keys. $\nu_2$ is obtained by dividing the  experimentally determined total number of 
collisions by $N/2$. }
\label{Fig_6}
\end{figure}
Whereas all of the results  so far are for systems containing 400 particles,
the dependence of the KS-entropy on the single-particle collision frequency $\nu_2$ is 
demonstrated in Fig. \ref{Fig_6} for a system with only $N=64$ disks (to reduce the
computational cost). This number is still large enough to be representative for large systems.
As was found for the maximum Lyapunov exponent,  the phase transition 
(near $\nu_2 \approx 20$) does not specifically show up in $h_{KS}$ when 
viewed as a function of $\nu_2$ instead of $\rho$. If $\rho$ approaches 
the close-packed density $\rho_0 = 1.1547$, both $\lambda_1$ and $h_{KS}$ diverge due to the
divergence of $\nu_2$. For very low densities and smooth hard disks ($\kappa = 0$), 
a kinetic-theory based density expansion for the KS-entropy similar to that in 
Eq. \ref{vanZon} becomes \cite{vB_1997,vZ1} 
\begin{equation}
h_{KS} / N =   A' \nu_2 \left[ - \ln \rho + B' + {\cal O}( \rho ) \right].
\label{hks_est}
\end{equation}
Estimates for the constants $A',B'$ have been obtained by 
van Zon et al. \cite{vZ1} and most recently by de Wijn \cite{Astrid}, $A'= 0.5, B'=1.47 \pm 0.11$,
which describe the numerical data well for  $\rho < 10^{-3}$.
Again we expect a similar representation to hold also for $\kappa > 0$. Still, it seems surprising that
the KS-entropy is reduced so much by the introduction of an internal degree of freedom (rotation),
which effectively acts as energy storage in between collisions.
  
   An interesting step-like structure is observed for very small $\kappa$ in
Fig. \ref{Fig_2tb} and in the upper panel of Fig. \ref{Fig_4tb}. These steps are a remnant of the
degeneracy of  exponents due to the existence of Lyapunov modes for {\em smooth} ($\kappa \equiv 0$) 
hard disk systems \cite{PH_2000,EFPZ_2005}. Lyapunov modes are periodic spatial perturbations
associated with the small positive exponents with indices $l < l_0 = 2N-3$ (and  with the
conjugately paired negative exponents). The corresponding perturbation vectors may be represented as periodic vector fields coherently spread out over the simulation box and with well defined wave vectors.
They may be understood as Goldstone modes of a system with continuous symmetries \cite{Goldstone}
- translation invariance in space and time - which give rise to  conservation of energy and linear momentum \cite{Scheck} and in addition to the six vanishing exponents
\cite{gaspard_1998}.  Note that angular momentum is globally not conserved according 
to the periodic boundaries and does not contribute. Fig. \ref{Fig_2tb} shows that the exponent degeneracy and, hence the Lyapunov modes are still rather well developed for $\kappa = 0.004$ corresponding to a moment of inertia  $I = 0.001$. This is independent of the density as is shown in the upper panel of  Fig. \ref{Fig_4tb}. However, if $\kappa$ is increased,  the steps quickly disappear and Lyapunov modes do not seem to exist any more. This is most clearly demonstrated in Fig. \ref{Fig_2tb}.
 It is not clear to us why this happens in view of the fact that modes are readily found
 for two-dimensional hard dumbbell fluids. Fourier transformation techniques will be
 required to settle this point.  
 
 For rough hard disks the angular velocity subspace of the full phase space 
 has $N$ dimensions and contributes $N$ exponents to the full spectrum. 
 For small $\kappa$ these exponents are different from zero, but small. Half of them
 belong to the positive branch (to which we restrict ourselves without loss of generality) 
 and are located in the index interval  $2N-2 \le l \le (5N/2)-3$, sandwiched between 
 the translation-dominated regime, $ l \le 2N-3 $, and the three vanishing exponents 
 still attributed to the positive branch,  $ (5N/2) -2 \le l  \le 5N/2 $. We refer to this regime
 as rotation dominated. If $\kappa$ is increased, the exponents in this regime are increased,
and the exponents in the translation dominated  regime become smaller, until the spectrum
becomes very uniform as, for example,  in the lower panel of Fig. \ref{Fig_4tb}, and
the separation into translation and rotation dominated regimes becomes meaningless.
Such a system we call fully coupled. Translation and rotation contribute indistinguishably
to the mixing process in phase space.
 

\subsection*{Localization of tangent-space perturbations}


The maximum (minimum) Lyapunov exponent is the rate constant for the
fastest growth (decay) of a phase-space perturbation and is dominated by the
fastest dynamical events, binary collisions. It is not too surprising that the associated
tangent vector components are significantly different from zero for only a few
strongly-interacting particles at any instant of time. Thus, the respective perturbations 
are strongly localized in physical space. It has been shown that for both hard and soft
disk respective sphere systems the localization persists in the thermodynamic limit, 
such that the fraction of tangent-vector components contributing to the generation of $\lambda_1$ 
follows a power law $\propto N^{-\eta}, \eta > 0$, and
converges to zero for $N \to \infty$ \cite{MP_2002,PF_2002,FHPH_2004,forster_posch_2005}.   
The localization becomes gradually worse for larger indices $l > 1$, until it ceases to
exist and (almost) all particles collectively contribute to the coherent Lyapunov modes
mentioned in the previous section. Similar observations for spatially extended systems 
have been made by various authors \cite{Manneville,LR_1989,FMV_1991,TM_2003a,TM_2003b}. 

\begin{figure}[t]
\centering{
\vspace{-8mm}
\includegraphics[width=0.45\textwidth]{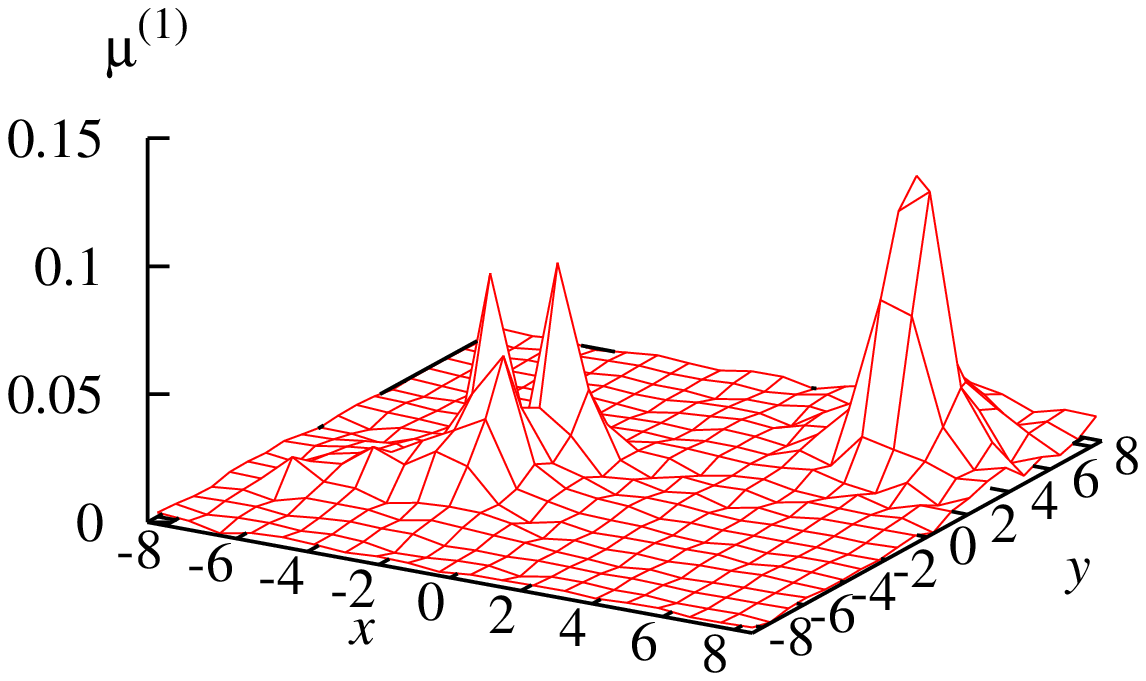}\\
\vspace{-18mm}
\includegraphics[width=0.45\textwidth]{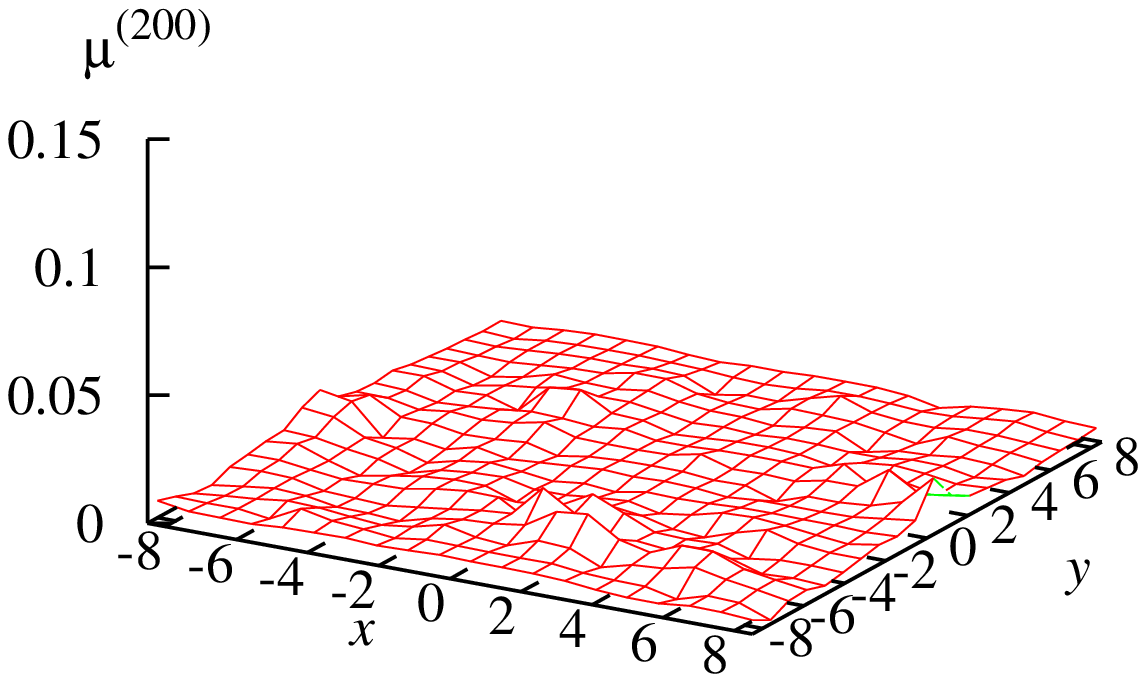} \\
\vspace{-5mm}
\includegraphics[width=0.45\textwidth]{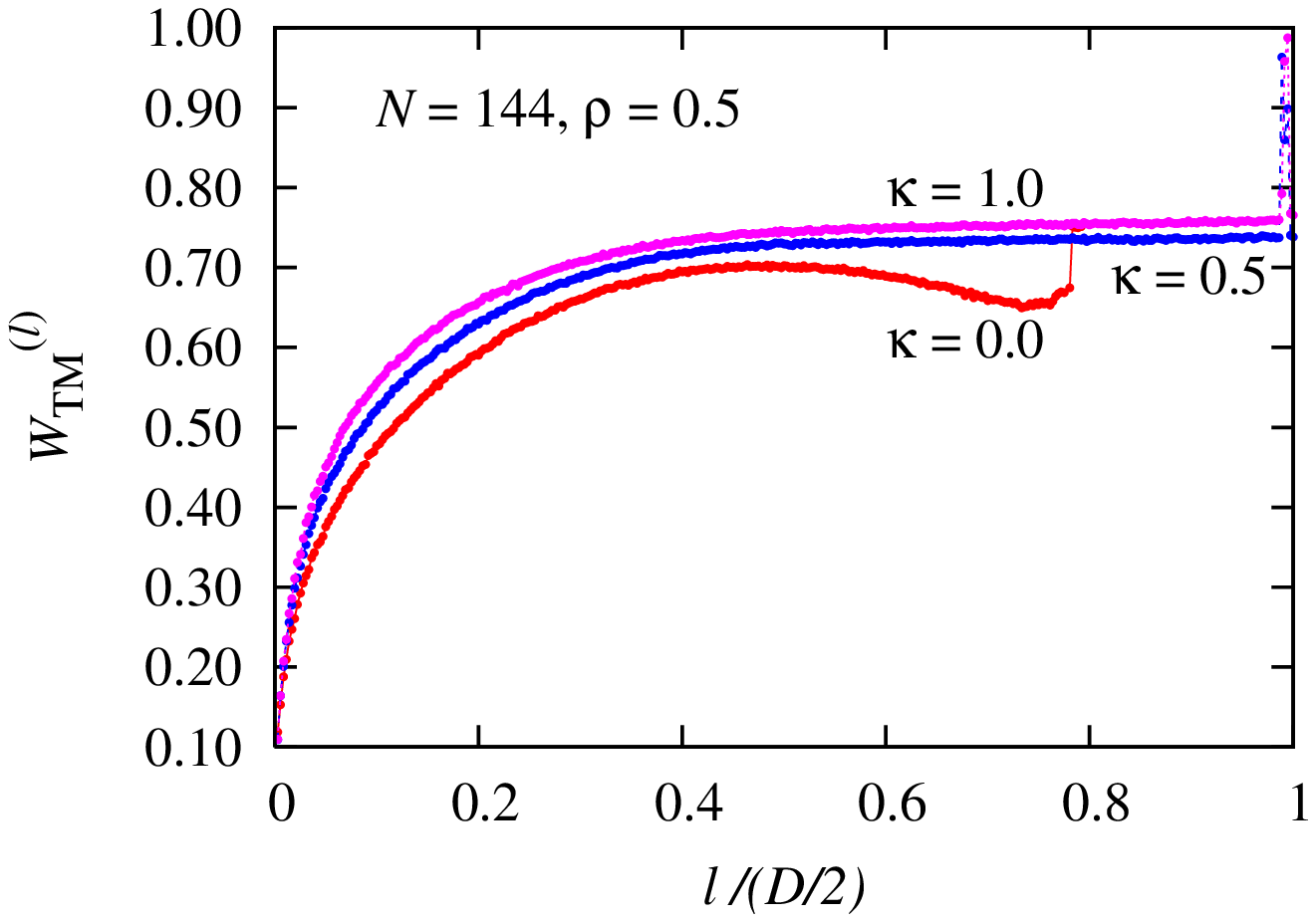}}
\caption{ (Color online) Top: Instantaneous view of the action probability $\mu_i^{(1)}$  for the 
maximally localized tangent vector, plotted as a scalar field
over the simulation box. For the grid interpolation a weight function $w(r) \propto r^{-2}$
is used, where $r$ is the distance of a particle to the grid point. The system consists 
of $N=144$ rough hard disks with $\kappa = 0.5$ at a density $\rho = 0.5$. \\
Middle: As in the top panel, but for $\mu_i^{(200)}$, which belongs to a delocalized vector
with normalized index $\bar{l} = 0.556$.\\
Bottom: Localization width $W_{TM}^{(l)}$ (see Eq. (\ref{locwidth}))  for $N=144$ disks with  
$\kappa$ indicated by the labels. The density $\rho = 0.5$. The reduced index $\bar{l} = l/(D/2)$ is used
on the abscissa, and $D = 5N$. }
\label{Fig_7}
\end{figure}
    To demonstrate the localization property of the rough hard-disk system, we define
the contribution of an individual  disk $i$ to the perturbation vector  ${\vec\delta \Gamma}^{(l)}$
belonging to $\lambda_l$ as the square of the projection of  ${\vec\delta \Gamma}^{(l)}$
onto the subspace of this disk,
 $$\mu_i^{(l)} =\Big( \vec{\delta q_i}^{(l)} \Big)^2 +\Big(\vec{\delta v_i}^{(l)}\Big)^2 +
  \Big(\delta \omega_i^{(l)}\Big)^2 .$$
Because ${\vec\delta \Gamma}^{(l)}$ is normalized
in the Gram-Schmidt step of the algorithm, one has $\sum_i^N \mu_i^{(l)} = 1$ for all $l$, and 
$\mu_i^{(l)}$ may be interpreted as a kind of
(normalized) action probability of $i$ for the perturbation $l$.  It should be noted that
for the definition of $\mu_i^{(l)}$ the Euclidean norm is used and that all localication measures
depend on this choice. Qualitatively, this is sufficient to demonstrate localization.

In the top panel of Fig. \ref{Fig_7}, $\mu_i^{(1)}$ for $l=1$ is plotted as a scalar 
field $\mu^{(1)}(\vec{q})$,
where the surface is interpolated over regular grid points covering the 
whole simulation box. There exist  one big and a few smaller competing active zones,
which move around randomly such that the system remains homogeneous
when viewed over a long time. This should be contrasted with the middle panel,
where an analogous plot (with the same scale) is shown for the delocalized
tangent vector with an index $l = 200$ $(\bar{l}= 0.556).$
For this comparison, the system contains 144 rough disks with $\kappa = 0.5$
at a density $\rho = 0.5.$

       A number of localization measures have been introduced to assess the localization of
 ${\vec\delta \Gamma}^{(l)}$,  not only for   $l=1$  \cite{MP_2002}   but for all $l$ 
 \cite{TM_2003a,TM_2003b}.  The most common is due to Taniguchi and Morriss \cite{TM_2003a},  
 who define a ''localization width''
 \begin{equation}
 W_{TM}^{(l)}  = \exp[S^{(l)}] /N ,
 \label{locwidth} 
 \end{equation}  
 which is based on the Shannon entropy for the ''probability''  distribution  $\mu_i^{(l)}$: 
 \begin{equation}
S^{(l)}  = \left\langle - \sum_{i=1}^N \mu_i^{(l)} \ln \mu_i^{(l)} \right\rangle .
\nonumber
 \end{equation}   
 Here, $\langle \cdots \rangle$ denotes a time average. $W_{TM}^{(l)}$
 is bounded according to $1/N \le W_{TM}^{(l)} \le 1$,  where the lower and upper bounds apply
 for complete localization and delocalization, respectively. In the bottom panel of Fig. \ref{Fig_7} 
we plot   $W_{TM}^{(l)}$ for 
the 	144-disk system used before. The value of $\kappa$ is indicated by the labels. 
This localization spectrum changes surprisingly little when $\kappa$ is increased from zero to one.
The only major difference is for the rotation-dominated regime. For the smooth disks,
$\kappa = 0$,  the points for $W_{TM}^{(l)}$ are irrelevant in this regime, $0.8 \le \bar{l} \le 1.2$, and are not shown. Note that  only data for the positive branch of the Lyapunov spectrum are shown, $\bar{l} \le 1$.
 
  Alternatively, an even simpler definition may be used, which involves the  Fermi entropy (sometimes also  referred to as the quadratic entropy), 
 \cite{Jumarie}
 \begin{equation}
        S_F^{(l)}  =  \left\langle \sum_{i=1}^N      \mu_i^{(l)} ( 1 - \mu_i^{(l)} ) \right\rangle.
  \label{Fermi}      
\end{equation}
It  has the desired property: $S_F^{(l)}$ vanishes, if only a single particle is responsible for
 the phase-space growth (extreme localization), it is $(N-1)/N \approx 1$, if all particles contribute identically (complete coherent delocalization), and it is
 in between otherwise. This measure might be particularly useful, whenever  localization is even more complete  than in the case presented here, but it distinguishes poorly between 
 very delocalized states..
 
At this point, a critical remark is in order. The localization spectrum in the bottom panel of
Fig. \ref{Fig_7} is shown for the positive branch of the Lyapunov spetrum only. It should
be completely symmetrical with respect to the conjugate negative branch, $S^{(l)} = S^{(D+1-l)}$,
due to the time reversible phase-space structure: a time reversal operation converts the stable manifold
into the unstable manifold and vice versa. For the smooth hard-disk case, $\kappa= 0$,
this symmetry is  observed with high numerical precision \cite{JvM_2005,BP_2009}. 
However, for $\kappa > 0$ the spectra are slightly asymmetric (not shown). 
The reason for this asymmetry is  subtle. The perturbation
vectors   ${\vec\delta \Gamma}^{(l)}$ we use in this work are ortho-normal. They span the 
correct subspaces of the tangent space required for the computation of the Lyapunov exponents
according to the standard algorithm \cite{Benettin, Shimada,EFPZ_2005}, but they
are not covariant: that means, they do not strictly follow the linearized dynamics in 
tangent space, but are regularly re-orthonormalized by the Gram-Schmidt procedure. As a 
consequence, they are not invariant under time reversal. The last property, however, is required
for a complete symmetry of the localization spectrum, such that the  
expanding vector ${\vec\delta \Gamma}^{(l)}$ in the time-forward direction becomes the
contracting vector  ${\vec\delta \Gamma}^{(D+1-l)}$  in the time-backward direction.
If proper covariant Lyapunov vectors 
\cite{Ginelli} are used instead of the Gram-Schmidt vectors, the symmetry is re-established
for $\kappa > 0$. Details will be communicated in a forthcoming publication \cite{BP_2009}.

%
\section*{Convergence times in tangent space}
%

   For the Lyapunov exponents to converge, the orthonormal set of tangent vectors
 needs to reach  its proper orientation in tangent space starting from an arbitrary
 initial orientation. The  convergence time varies with the number of particles and with the index 
 $l$. For smooth particle systems it was shown in Ref. \cite{DHP_2002} that the vector associated 
 with the maximum exponent aligns with a convergence time proportional to $N^{\alpha}$, 
 where $\alpha $ lies between 0.4 (for smooth hard disks) and 0.9 (for soft
 repulsive interaction potentials). For higher indices the convergence is an even slower
 collective phenomenon. In this section the methods of Ref. \cite{DHP_2002} are adapted
 to determine the system-size dependence of the convergence times of rough disks not only
 for $l = 1$ but for all tangent vectors.

     We consider $M$ randomly oriented orthonormal sets of tangent vectors,
$\Delta_m = \{ \vec{\delta\Gamma}_m^{(l)}\}, l=1,\cdots, D$, where $ m=1,\cdots,M$.
Each set  spans the full $D$-dimensional tangent space and acts as an  initial condition for the computation of  a full  Lyapunov spectrum for the same reference trajectory. All spectra eventually converge. Any two tangent vectors giving rise to the {\em same} Lyapunov exponent but belonging to
{\em different} initial sets $\Delta$ need to become parallel or anti-parallel in the course of time,
such that  their dot-product approaches $\pm 1$. To measure the convergence time for a given $l$, we average over all such possible products,
\begin{equation}
\chi_l(t) = \frac{2}{M (M-1)} \sum_{m=1}^{M-1} \sum_{m' = m+1}^{M}
\left| \vec{\delta \Gamma}_m^{(l)} \cdot \vec{\delta \Gamma}_{m'}^{(l)} \right| . 
\label{chi}     
\end{equation}
$\chi_l(t)$ increases with time from $\chi_0$ to unity, where  the initial value
($\chi_0 \approx 0.06$  for $N = 36$) is independent of $l$ due to the random orientation of
the sets $\Delta$ and converges to zero for $N \to \infty $.  The time for which $\chi_l(t)$ crosses a threshold $\Theta$  for the first time is taken as a measure of the convergence time $\tau_l$.
In the following  we take  $\Theta = 0.9$.  Any other choice for this threshold only results in 
times which differ by a constant factor. 

\begin{figure}[ht]
\center
\includegraphics[width=0.45\textwidth]{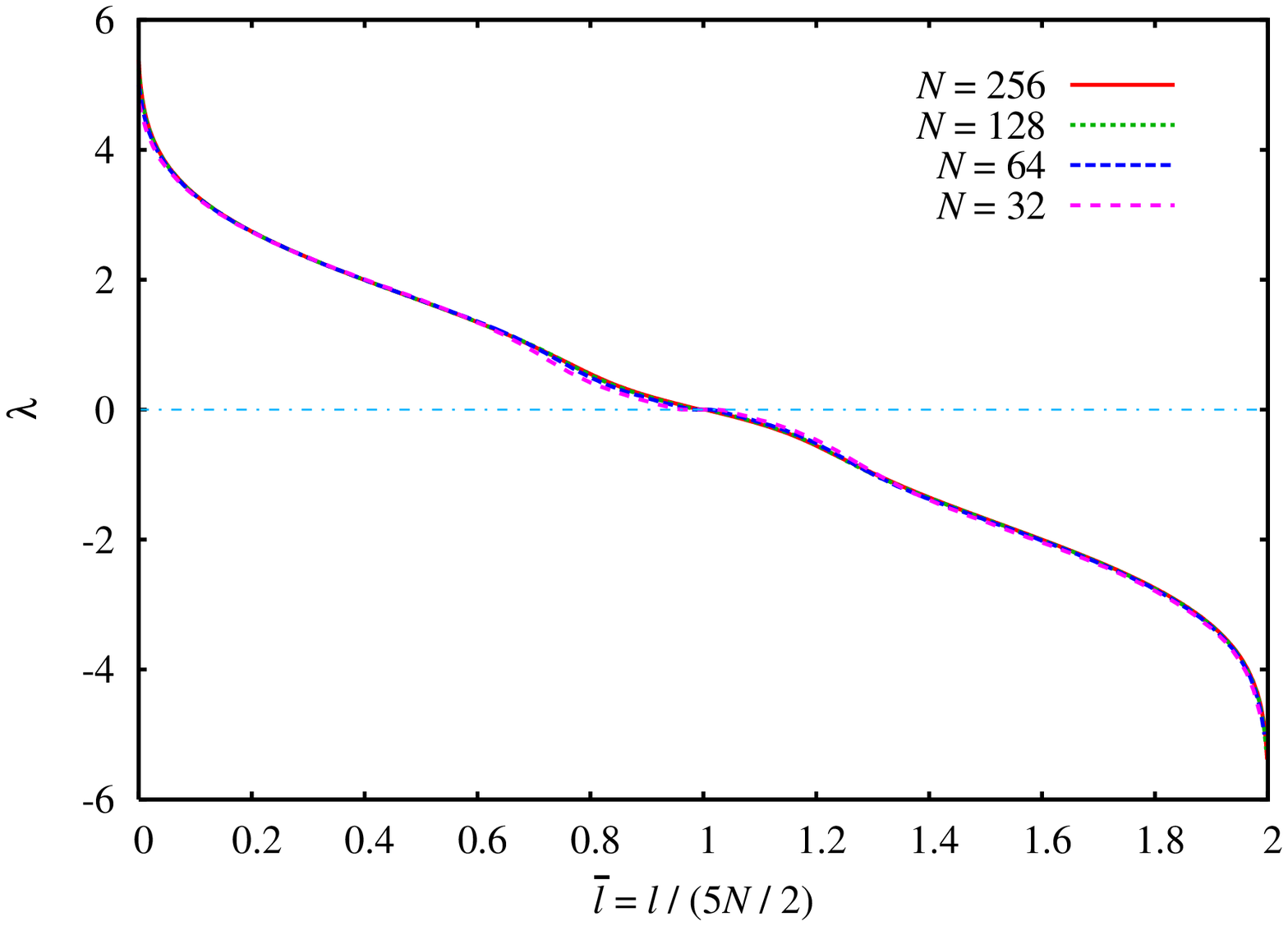}
\includegraphics[width=0.45\textwidth]{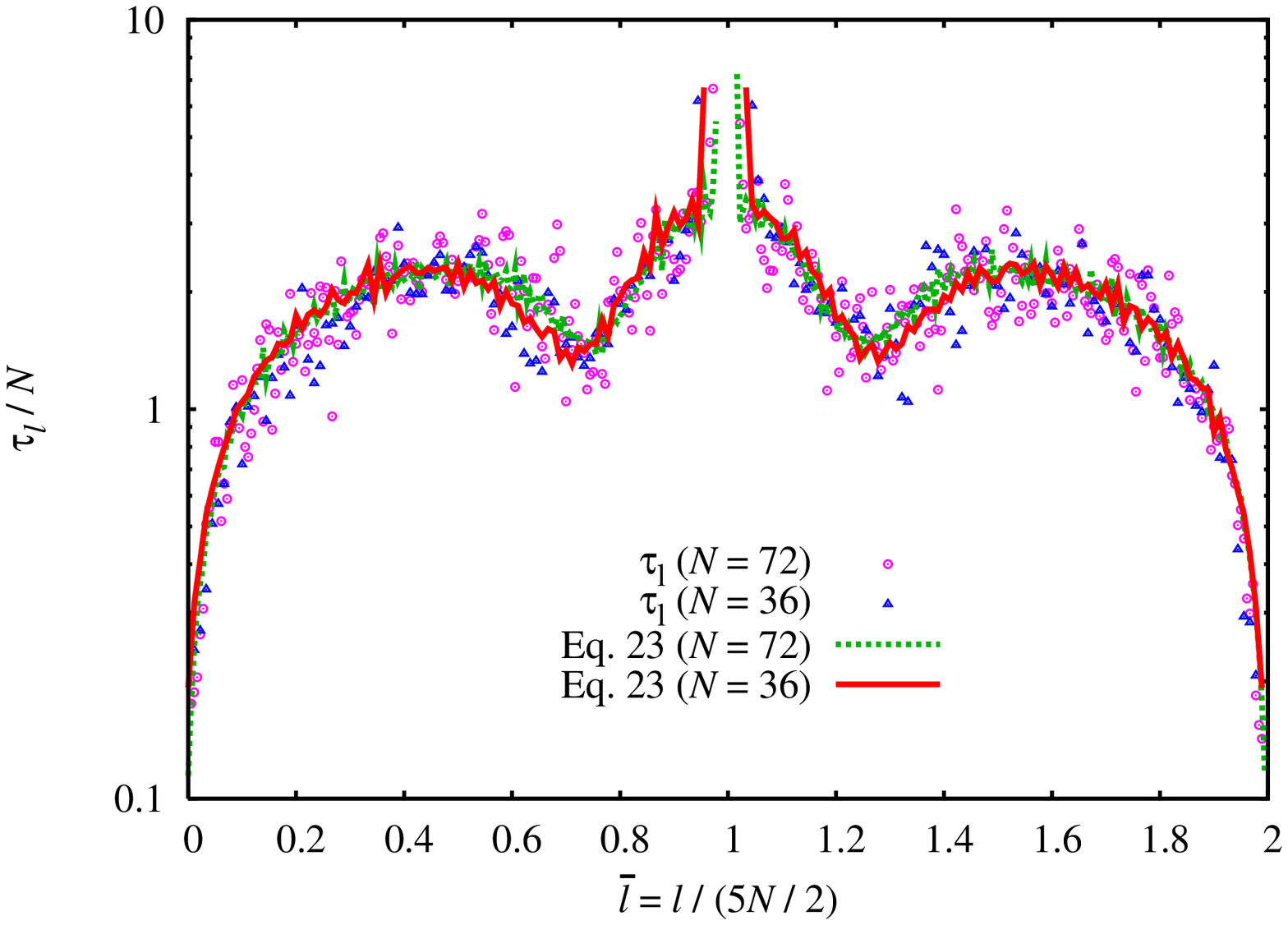}
\caption{(Color online) Top: Full Lyapunov spectra of rough hard disk systems for various 
system sizes $N$, plotted  as a 
function of the normalized  exponent $\bar{l} = l /(5N/2)$. All
systems have a density $\rho = 0.7$, and  $\kappa=0.4$.
The spectra quickly converge to a smooth limit  spectrum for $N \to \infty$. 
Bottom: Lyapunov vector convergence times, divided by $N$, for two system sizes, 
$N=36$ and $N=72$ as indicated by the keys. The points are direct simulation results,
the smooth lines were computed from the Lyapunov spectrum via Eq. (\ref{taushort}),
where $A=2.85$.}
\label{Fig_8}
\end{figure}

Before continuing the discussion, we note that
the Lyapunov spectrum exists in the thermodynamic limit. This has been shown for smooth hard 
disks and hard spheres \cite{DPH_1996} and means that for $N \to \infty$, at constant density,
the Lyapunov spectra quickly converge to a limiting curve when plotted as a function
of the reduced index $\bar{l} = l/(D/2)$. For the rough hard disks this is demonstrated in
the upper panel of Fig. \ref{Fig_8}. The spectra there are for 16 to 256 particles
at a density $\rho = 0.7$, and for $\kappa = 0.4$. Full spectra with their positive 
and negative branches are shown, which are  related by the conjugate pairing symmetry 
as was mentioned in Sec. \ref{section_model}. 
Thus, the derivative of the spectrum with respect to the normalized index exists in that limit,
\begin{equation}
\frac{d \lambda( \bar{l} )}{ d \bar{l} } = \lim_{N \rightarrow \infty}
\frac{\lambda_{\bar{l} + \Delta\bar{l}} - \lambda_{\bar{l}}}{ \Delta \bar{l}}.
\label{derivative}
\end{equation}

It has been argued in Ref. \cite{DHP_2002} that  for $\kappa = 0$ the decay time for the 
correlation  function  $\chi_1(t)$ concerning the maximum exponent  
is determined by $1/(\lambda_1 - \lambda_2)$ and, hence, by
the inverse ''slope'' of the spectrum at $l=1$. These arguments also apply to all the other 
exponents such that one expects
\begin{equation}
\tau_l = \frac{A}{ |\lambda_{l+1} - \lambda_l |}
\label{taushort}
\end{equation}
to hold, where $A$ is a fitting parameter which, for the choice $\Theta = 0.9$, 
becomes 2.85.  
Rewriting this expression in terms of the reduced index $\bar{l} =  l / (5N/2) $ gives
\begin{equation}
\tau_{l} = \frac{A} {\left|  \Delta \lambda(\bar{l}) / \Delta \bar{l} \right| \Delta \bar{l} }
       \approx  \frac{ 5N A }{2} \left| \frac{ d \lambda(\bar{l}) }{ d \bar{l}} \right|^{-1} ,
\label{tau}
\end{equation}
where we have replaced the finite differences by the respective differentials. Since the
limiting spectral slope, Eq. (\ref{derivative}), is independent of $N$, the convergence 
time for any $l$ is expected to be proportional to the particle number $N$. 

Our results for $\tau_l/N$ are depicted in the lower panel of Figure
\ref{Fig_8}, where experimental results for  $N=36$ and $N=72$ rough disks are shown 
by the points. Both systems have a density $\rho = 0.7$, and  $\kappa=0.4$.
Clearly the points for different $N$ collapse onto a single curve proving the proportionality of
$\tau_l$ to $N$. If $\tau_l$ is computed from the slope of the
 Lyapunov spectrum according to Eqs. (\ref{taushort}) or (\ref{tau}), the smooth
 lines are obtained. Their agreement with the simulation points supports our
 assertion in Eq. (\ref{taushort}).
 
 The symmetry obviously exhibited by the convergence time, $\tau_l = \tau_{D+1 - l}$,
 is surprising in view of the fact that the algorithm treats successive vectors 
 successively:  The orientation of the second vector is affected by that of the first,
 that of the third vector by that of the first and second, and so on. 

\section{Conclusions}
\label{conclude}

In this paper we investigate rough hard-disk systems, arguably
the most simple models of a molecular fluid with translational
and rotational degrees of freedom.  The rotation of the particles may be
viewed as a mechanism to store  internal energy, which is returned to 
the translational degrees of motion with some delay.  We compute Lyapunov spectra and
study the effect of rotation-translation coupling on the 
dynamical stability of such systems.

If the moment of inertia $I$ of the disks vanishes,
the translational dynamics is completely decoupled from the rotational 
degrees of freedom and the results for the smooth
hard-disk system are reproduced. If $I$ respective the more relevant coupling parameter
$\kappa = 4 I$ is increased, the Lyapunov spectrum changes drastically with
the rotation-dominated parts of the spectrum being gradually filled in,
until a separation into translation- and rotation-dominated parts becomes 
meaningless.

The maximum exponent, $\lambda_1$, which is taken as an indicator for dynamical
chaos, increases with increasing $\kappa$ for large enough densities ($\rho > 0.7$), but decreases
for smaller densities. At the same time, the Kolmogorov-Sinai entropy $h_{KS}$ always
decreases. The latter, which is the sum of all positive exponents, gives the rate
of mixing in phase space, which becomes less and less effective the more  important the rotation
is for the dynamics. We encounter the unexpected situation that for large densities 
dynamical chaos may increases with $\kappa$ whereas at the same time phase-space  mixing  
takes longer. This should be contrasted to the behavior of a system of hard dumbbells
\cite{MP_2002}. For a uniform mass distribution of the dumbbells (corresponding to 
$\kappa = 0.5$ for the rough disks), both $\lambda_1$ and $h_{KS}$ increase with
the molecular anisotropy, and the mixing time decreases. From this point of view, the
rough disk model seems artificial.

Another surprise is the seeming lack of Lyapunov modes for the rough disks with 
non-vanishing $\kappa$,  given the fact that modes were readily found for hard-dumbbell 
systems \cite{MP_2002}. As for soft interaction potential systems, Fourier transformation
methods may still give evidence for modes.
This point deserves further investigation. However, the localization in physical space of the perturbation 
vectors associated with the maximum exponent is as expected.

The localization spectrum shown in the bottom panel of  Fig. \ref{Fig_7} is an 
application of a projection of the tangent vectors onto the phase space of
individual disks. Due to the time-reversal symmetry of the evolution equations, 
such projections should show definite symmetries with respect to the positive
and negative (not included in Fig. \ref{Fig_7}) branches of the Lyapunov spectrum.
However, more often than not, these symmetries are numerically not recovered by the
classical algorithm. The explanation lies in the fact that Gram-Schmidt-orthonormalized
tangent vectors span the proper subspaces for the computation of the exponents, but
are not covariant with the tangent flow \cite{Ginelli}. If covariant vectors are used,
these spurious asymmetries disappear \cite{BP_2009}.  

Up to now little is known about the mechanism governing  the convergence of 
tangent vectors towards their proper directions. For the rough hard disk systems 
it is shown that the convergence time for all $l$ vary linearly with the system size, $N$. 
Furthermore,  they are related to the slope of the spectrum at a particular $l$. This
view is suggested by the existence of the thermodynamic limit of the specrum.

 \section{Acknowledgments} 
We thank Hadrien Bosetti for fruitful remarks,  and Christoph Dellago and William G. Hoover for interesting discussions. We gratefully acknowledge support from the Austrian Science 
Foundation (FWF) under grant No. P18798-N20.

\end{document}